\documentclass[a4paper]{article}
\usepackage{amssymb}

\usepackage{graphicx}
\usepackage{amsmath}

\begin{document}

\title{Simultaneous amplitude and phase damping of a kind of Gaussian states and
their separability}
\author{Xiao-yu Chen \\
{\small {Lab. of Quantum Information, China Institute of Metrology,
Hangzhou, 310018, China}}}
\date{}
\maketitle

\begin{abstract}
We give out the time evolution solution of simultaneous amplitude and phase
damping for any continuous variable state. For the simultaneous amplitude
and phase damping of a wide class of two- mode entangled Gaussian states,
two analytical conditions of the separability are given. One is the
sufficient condition of separability. The other is the condition of PPT
separability where the Peres-Horodecki criterion is applied. Between the two
conditions there may exist bound entanglement. The simplest example is the
simultaneous amplitude and phase damping of a two-mode squeezed vacuum
state. The damped state is non-Gaussian.

PACS 03.65.Ud 03.65.Yz 03.67.Mn 42.50.Dv
\end{abstract}

\section{Introduction}

Quantum entanglement or inseparability plays a major role in all branches of
quantum information and quantum computation. Peres\cite{Peres} proposed a
criterion for checking the inseparability of a state by introducing the
partial transpose operation. This condition is necessary and sufficient for
some lower dimensional discrete bipartite systems but is no longer
sufficient for higher dimensions\cite{Horodecki}. Despite many studies on
the discrete states, much attentions have been paid to the continuous
variable states \cite{Braunstein}. Recently, quantum teleportation of
coherent states has been experimentally realized by exploiting a two-mode
squeezed vacuum state as an entanglement resource\cite{Furusawa}. Due to the
decohence of the environment, a pure entanglement state will become mixed.
Thus it is important to know if a given bipartite continuous variable state
is entangled or not. The decoherence may be caused by coupling to the
thermal noise of the environment, amplitude damping, quantum dissipation and
phase damping. Besides the phase damping, the other three types of
decoherence preserve gaussian property of the state, and a two-mode squeezed
vacuum state will evolve to a two mode gaussian mixed state. For the
separability of two mode gaussian state, the positivity of the partially
transposed state is necessary and sufficient \cite{Duan}\cite{Simon}\cite
{Wang}. However, a gaussian state will evolve to a non-gaussian one by phase
damping, and the case of a two mode squeezed vacuum state under the only
decoherence of phase damping was perfectly solved\cite{Hiroshima}. In real
experiments, the general situation which should be taken into account is the
coexistence of noise, amplitude and phase damping. Theoretically, the
separability and entanglement of non-gaussian state are seldom investigated,
here we provide an example.

\section{Time Evolution of Characteristic Function}

Considering the simultaneous damping, the density matrix obeys the following
master equation in the interaction picture $\frac{d\rho }{dt}=(\mathcal{L}_1%
\mathcal{+L}_2)\rho .$ Where $\mathcal{L}_1$ is the amplitude damping part
\begin{equation}
\mathcal{L}_1\rho =\sum_i\frac{\Gamma _i}2[(\overline{n}_i+1)(2a_i\rho
a_i^{+}-a_i^{+}a_i\rho -\rho a_i^{+}a_i)+\overline{n}_i(2a_i^{+}\rho
a_i-a_ia_i^{+}\rho -\rho a_ia_i^{+})],
\end{equation}
with $\overline{n}$ the average photon number of the thermal environment.
And $\mathcal{L}_2$ is the phase damping part (e.g. \cite{Hiroshima}),
\begin{equation}
\mathcal{L}_2=\sum_i\frac{\gamma _i}2[2a_i^{+}a_i\rho
a_i^{+}a_i-(a_i^{+}a_i)^2\rho -\rho (a_i^{+}a_i)^2].
\end{equation}
The state can be equivalently specified by its characteristic function.
Every operator $\mathcal{A}\in \mathcal{B(H)}$ is completely determined by
its characteristic function $\chi _{\mathcal{A}}:=tr[\mathcal{AD}(\mu )]$
\cite{Petz}, where $\mathcal{D}(\mu )=\exp (\mu a^{+}-\mu ^{*}a)$ is the
displacement operator, with $\mu =[\mu _1,\mu _2,\cdots ,\mu _s]^T$ $%
,a=[a_1,a_2,\cdots ,a_s]$ and the total number of modes is $s.$ It follows
that $\mathcal{A}$ may be written in terms of $\chi _{\mathcal{A}}$ as \cite
{Perelomov}: $\mathcal{A}=\int [\prod_i\frac{d^2\mu _i}\pi ]\chi _{\mathcal{A%
}}(\mu )\mathcal{D}(-\mu ).$ The density matrix $\rho $ can be expressed
with its characteristic function $\chi $. The amplitude damping equation of $%
\chi $ and its solution are well known, they are $\frac{\partial \chi }{%
\partial t}=-\frac 12\sum_i\Gamma _i(\left| \mu _i\right| \frac{\partial
\chi }{\partial \left| \mu _i\right| }+\left| \mu _i\right| ^2\chi ),$ $\chi
(\mu ,t)=\chi (\mu _ie^{-\frac{\Gamma _it}2},0)\exp [-\sum_i(\overline{n}%
_i+\frac 12)(1-e^{-\Gamma _it})\left| \mu _i\right| ^2]$. We now give out
the phase damping equation of $\chi $, it will be
\begin{equation}
\frac{\partial \chi }{\partial t}=\frac 12\sum_i\gamma _i\frac{\partial
^2\chi }{\partial \theta _i^2}
\end{equation}
if we denote $\mu _i$ as $\left| \mu _i\right| e^{i\theta _i}$. We can see
that with the characteristic function the amplitude damping equation is
described by the amplitude of the parameter $\mu _i,$ the phase damping
equation is described by the phase of the parameter $\mu _i$. The solution
to the phase equation of $\chi $ then will be $\chi \left( \mu ,\mu
^{*},t\right) =\prod_i\left( 2\pi \gamma _it\right) ^{-1/2}\int \exp (-\sum_i%
\frac{x_i^2}{2\gamma _it})$ $\chi \left( \mu e^{ix},\mu ^{*}e^{-ix},0\right)
dx,$ where $\mu e^{ix}$ is the abbreviation of $[\mu _1e^{ix_1},\mu
_2e^{ix_2},\cdots ,\mu _se^{ix_s}]$.The simultaneous amplitude and phase
damping to any initial characteristic function then will be
\begin{equation}
\chi \left( \mu ,\mu ^{*},t\right) =\prod_i\left( 2\pi \gamma _it\right)
^{-1/2}\int \exp \sum_i[-\frac{x_i^2}{2\gamma _it}+(\overline{n}_i+\frac
12)(1-e^{-\Gamma _it})\left| \mu _i\right| ^2]\chi \left( \mu _ie^{-\frac{%
\Gamma _it}2+ix_i},\mu _i^{*}e^{-\frac{\Gamma _it}2-ix_i},0\right) dx.
\end{equation}
The density matrix then can be obtained as well by making use of operator
integral.

We will concentrate on the simultaneous amplitude and phase damping of
two-mode $x-p$ symmetric Gaussain state\cite{Jiang} whose characteristic
function is $\chi (\mu ,0)=\exp [-(A_{10}\left| \mu _1\right|
^2+A_{20}\left| \mu _2\right| ^2)+B_0\mu _1\mu _2+B_0^{*}\mu _1^{*}\mu
_2^{*}].$ . The time evolution of $\chi $ is
\begin{equation}
\chi \left( \mu ,\mu ^{*},t\right) =\int_x\exp [-(A_1^{\prime }\left| \mu
_1\right| ^2+A_2^{\prime }\left| \mu _2\right| ^2)+B\mu _1\mu
_2e^{ix}+B^{*}\mu _1^{*}\mu _2^{*}e^{-ix})].
\end{equation}
Where $A_i^{\prime }=e^{-\Gamma _it}A_{i0}+(\overline{n}_i+\frac
12)(1-e^{-\Gamma _it}),$ $B=e^{-\frac 12(\Gamma _1+\Gamma _2)t}B_0,\overline{%
\gamma }=\frac 12(\gamma _1+\gamma _2)$. The integral on $x$ is one fold,
for simplicity, we denote $\frac 1{\sqrt{4\pi \overline{\gamma }t}}\int \exp
[-\frac{x^2}{4\overline{\gamma }t}]f(x)dx$ as $\int_xf(x).$ The $x-p$
symmetric Gaussain state set is a quite large set. It contains two-mode
squeezed vacuum state $\left| \Psi \right\rangle =\frac 1{\cosh
r}\sum_n(\tanh r)^n\left| n,n\right\rangle $ ( $r$ is the squeezing
parameter) and two-mode squeezed thermal state \cite{Chen} as its special
cases, with $A_{10}=A_{20}=\frac 12\cosh 2r$, $B_0=\frac 12\sinh 2r$ and $%
A_{10}=A_{20}=(n_0+\frac 12)\cosh 2r$, $B_0=(n_0+\frac 12)\sinh 2r$
respectively.

\section{PPT separability}

For the sake of simplicity in description, let us firstly consider the
situation of $A_1^{\prime }=A_2^{\prime }=A^{\prime },B=B^{*}.$ The general
case will be obtained straightforward and be described at the end of this
section. The first question is that if the state after damping is entangled
or not. Then how much is the entanglement left? The necessary condition of a
bipartite state being entangled is that the partial transpose of the density
operator is not positive definite \cite{Peres}. The partial transpose
operation changes the characteristic function in the fashion of : $\chi
\left( \mu _1,\mu _2\right) \Longrightarrow $ $\chi \left( \mu _1,-\mu
_2^{*}\right) =\chi ^{PT}\left( \mu _1,\mu _2\right) .$ For the separability
of a non-gaussian bipartite state, a necessary condition was proposed by
Simon\cite{Simon} in terms of the second moment of the state. In the
original literature canonical operators were used, here we use creation and
annihilation operators instead. The necessary condition comes from the
non-negativity of $\rho ^{PT}$ and the commutation relations. For any $%
Q=\eta \eta ^{+}$ with $\eta =c_1a_1+c_2a_2+c_3a_1^{+}+c_4a_2^{+}$ of every
set of complex coefficients $c_i$, one has $\left\langle Q\right\rangle
=tr(Q $ $\rho ^{PT})\geq 0,$ Hence the second moment matrix of $\rho ^{PT}$
should be semi-positive definite. The second moment such as $tr(\rho
^{PT}a_i^{+}a_j)$ can be obtained from the second derivative of $\chi $ with
respect to $\mu _i$ and $\mu _j^{*}$ , we have
\begin{equation}
A-1\geq Be^{-\gamma t}  \label{wave1}
\end{equation}
Where $A=A^{\prime }+\frac 12$ . Other necessary conditions may come from
when $\eta $ is the linear combination of higher power of the creation and
annihilation operators, and they may be tighter than Ineq.(\ref{wave1}). And
this is really the case.

We will find a tighter condition by exploring the negative eigenvalues of
the partial transpose of the density operator. The eigenequation of $\rho
^{PT}$ can be simplified as the eigenequations of a serials of matrices (see
Appendix).
\begin{equation}
M_{ln}^{(m)}=(A^2-B^2)^{-1}C^m\sum_k\binom{m-n}{l-k}\binom nk\left( \frac
DC\right) ^{l+n-2k}e^{-\gamma t(n-l)^2},
\end{equation}
where $C=1-\frac A{A^2-B^2},$ $D=\frac B{A^2-B^2}.$ When $m=0$, one has the
first eigenvalue $\lambda ^{(0)}=(A^2-B^2)^{-1}$ which is always positive.
The matrix $M^{(m)}$ possesses the symmetry of $%
M_{ln}^{(m)}=M_{m-l,m-n}^{(m)}$ , so that it can be reduced, and we get more
analytical solutions. The negative eigenvalues may appear at $\lambda
_0^{(1)}=(A^2-B^2)^{-1}(C-De^{-\gamma t}),$ $\lambda
_0^{(2)}=(A^2-B^2)^{-1}(C^2-D^2e^{-4\gamma t})$ and so on. Hence one of the
necessary conditions of the non-negativity of $\rho ^{PT}(t)$, so that the
necessary condition of a damped state $\rho (t)$ is PPT separable is that
\begin{equation}
C\geq De^{-\gamma t}.  \label{wave2}
\end{equation}

We will prove that this condition is also sufficient for PPT separability.We
turn to the detail properties of matrix $M^{(m)}$. The necessary condition
of separability comes from $M^{(2)}$ is $C\geq De^{-2d}$, this is a trivial
result compared with Ineq.(\ref{wave2}) We have checked other solvable
eigenvalues for necessary condition of separability. They are also trivial
compared with Ineq.(\ref{wave2}). It may be anticipated that separable
conditions come from all other $M^{(m)}$ are weaker than Ineq.(\ref{wave2}),
that is Ineq.(\ref{wave2}) is also a sufficient condition of PPT
separability. To prove this, we just need to consider the PPT separability
at the case of $C=De^{-\gamma t}$. Because if a state corresponding to $%
C=De^{-\gamma t}$ is PPT separable, then another state with stronger phase
damping (with increasing $\gamma $ while preserving all other parameters
unchanged) is definitely PPT separable, for this stronger phase damping
state we have $C>De^{-\gamma t}$ and it is PPT separable. Our proof of the
PPT separability of the state at $C=De^{-\gamma t}$ is not a most general
proof. We can only prove the non-negativity of $M^{(m)}$ by algebraic
programming up to $m=17$ at the case of $C=De^{-\gamma t}$. Denote $%
M_{ln}^{(m)}=(A^2-B^2)^{-1}C^mN_{ln}^{(m)},$ and let $N^{(m,j)}$ (with its
elements $N_{ln}^{(m,j)}$ , $0\leq l,n\leq j$) be the $j$-th main submatrix
of $N^{(m)}$, then to prove the non-negativity of $M^{(m)}$ is to prove that
the determinants of all $N^{(m,j)}$ are not negative. The algebraic
programming gives $\det N^{(m,j)}=d^{-p}(d-1)^{j(j+1)/2}P^{(m,j)}(d)\geq 0,$
where $d=D/C=e^{\gamma t}\geq 1,$ and $P^{(m,j)}(d)$ is a polynomial of $d$
with all its coefficients being positive integer, $p$ is some integer rely
on $j$. The algebraic programming runs for all $m\leq 17$ and proves the
non-negativity of $M^{(m)}$. We suggest that $M^{(m)}$ is also non-negative
for $m>17$, but this is not verified because of the computing time of the
algebraic programming.

Another direct way of proving comes from perturbation theory. Firstly $%
M^{(m)}$ can always be symmetrized. The zero order matrix is $%
M_{(0)}^{(m)}=M^{(m)}(\gamma t=0)$ which is just the case of gaussian state,
and all its eigenvalues and eigenvectors are well known. So that the first
order and second order perturbation of the eigenvalues of $M^{(m)}$ can be
obtained. A more concise way to obtain the perturbation result is as
follows: In the eigenequation of characteristic function, if we use $%
\left\langle \alpha \right| \left. \Phi \right\rangle =\exp (-\frac 12\left|
\alpha \right| ^2)\sum_{n=0}^mc_n^{\prime (m)}(\alpha _1^{*}+\alpha
_2^{*})^{m-n}(\alpha _1^{*}-\alpha _2^{*})^n$ as a test wave function, we
then get a matrix (see Appendix)
\begin{eqnarray}
M_{ln}^{\prime (m)} &=&\frac 1{(A^2-B^2)\sqrt{4\pi \gamma t}}\int dx\exp [-%
\frac{x^2}{4\gamma t}]\sum_k\binom{m-n}{l-k}\binom nk  \label{wave7} \\
&&(C+D\cos x)^{m-n-l+k}(C-D\cos x)^k(-1)^{l-k}(iD\sin x)^{l+n-2k}  \nonumber
\end{eqnarray}
which is a linear transformation of $M^{(m)}$ and has the same eigenvalues.
The zero order of $M^{\prime (m)}$ is the matrix $M_{(0)}^{\prime
(m)}=M^{\prime (m)}(\gamma t=0)$ which is a diagonal matrix. Hence
eigenvalues up to the first order perturbation of $M^{\prime (m)}$ are
simply $M_{nn}^{\prime (m)}$. When $\gamma t$ is quite small, $%
M_{nn}^{\prime (m)}$ can be approximated as $M_{nn}^{\prime (m)}\approx
(A^2-B^2)^{-1}(C+D)^{m-n}(C-D)^nf(m,n)$ with
\begin{equation}
f(m,n)=1-\gamma t[\frac D{D+C}(m-n)+\frac D{D-C}n+\frac{2D^2}{(D+C)(D-C)}%
(m-n)n].
\end{equation}
For odd $n$, the $n-th$ eigenvalue of the zero order approximation is
negative. If $f(m,n)$ is also negative, then the $n-th$ eigenvalue of the
first order approximation becomes positive. For given $\gamma t$ and $D/C$
we can always find sufficient large $m$ and proper $n$ so that $f(m,n)$ is
negative, hence the eigenvalues with large $m$ become positive faster than
that with small $m$ under phase damping. We need to know at what condition
all original negative eigenvalues become positive or zero. It is easy to
obtained that when $m=n=1$, $f(m,n)$ reaches its maximum. Hence when $%
f(1,1)=1-$ $\gamma t\frac D{D-C}\leq 0,$ that is
\begin{equation}
C\geq D(1-\gamma t),  \label{wave5}
\end{equation}
all other odd $n$ negative eigenvalue become positive. Ineq.(\ref{wave5}) is
the first order approximation of Ineq.(\ref{wave2}). Hence at the sense of
first order approximation Ineq.(\ref{wave2}) is sufficient for a state to be
PPT\ separable.

As a by product, we can use matrices $M^{(m)}$ to calculate the logarithmic
negativity which is an entanglement measure itself \cite{Vidal} and provides
an upper bound to the distillable entanglement\cite{Audenaert}. It can be
seen in the figure when $t\geq t_2,$the logarithmic negativity is zero,
while it is positive when $t<t_2$. For sufficiently small $\gamma t$, the
negativity of the state can be estimated. It is the absolute of the
summation of all eigenvalues with odd $n$. The result will be $\mathcal{N}%
(\rho )\approx \frac{D-C}{1+C-D}-\gamma t\frac{D(1-C+D)}{(1-C-D)(1+C-D)^2}=%
\frac{1-A+B}{2(A-B)-1}-\gamma t\frac B{[2(A-B)-1]^2}.$

One of the most important quantities of a state $\rho $ is its entropy $%
S(\rho )=-Tr\rho \log \rho $. The entropy of our damped state can be
obtained by solving the characteristic function eigenequation which is $\int
\frac{d^4\mu }{\pi ^2}\frac{d^4\alpha }{\pi ^2}\chi (\mu ,\mu
^{*},t)\left\langle \beta \right| D(-\mu )$ $\left| \alpha \right\rangle
\left\langle \alpha \right| \left. \Phi \right\rangle =\lambda \left\langle
\beta \right| \left. \Phi \right\rangle .$ After integrals on $\mu $, $%
\alpha $ and $x$, then compare the coefficients of each $\beta ^{*}$ item of
the two side, one can get a series of matrices $L^{(m)}$ whose eigenvalues
are that of the damped state $\rho $ and
\begin{equation}
L_{ln}^{(m)}=\frac{C^m}{A^2-B^2}\sum_k\binom{m+n}{n-k}\binom
lkC^{2k}(-D)^{l+n-2k}e^{-\gamma t(n-l)^2}.
\end{equation}
The entropy of the state $\rho $ will be $S(\rho )=-TrL^{(0)}\log
(L^{(0)})-2\sum_{m=1}^\infty TrL^{(m)}\log (L^{(m)}).$ The reduced state of $%
\rho $ is $\rho _1=Tr_2\rho $ with its characteristic function $\chi _1(\mu
_1,t)=\chi (\mu _1,0,t)=\exp [-A^{\prime }\left| \mu \right| ^2]$, hence its
entropy is $S(\rho _1)=A\log A-(A-1)\log (A-1).$ It has been proven that the
coherent information $I^i=\max (0,S(\rho _i)-S(\rho ))$ provides lower bound
of distillable entanglement of the state\cite{Devetak}. The coherent
information is calculated and plotted in the figure. At time $t_0$ the
coherent information turns to zero. In the figure we have $t_0<t_1$, but for
other parameters we may have $t_0>t_1$.

\begin{figure}[tbp]
\includegraphics[height=3in]{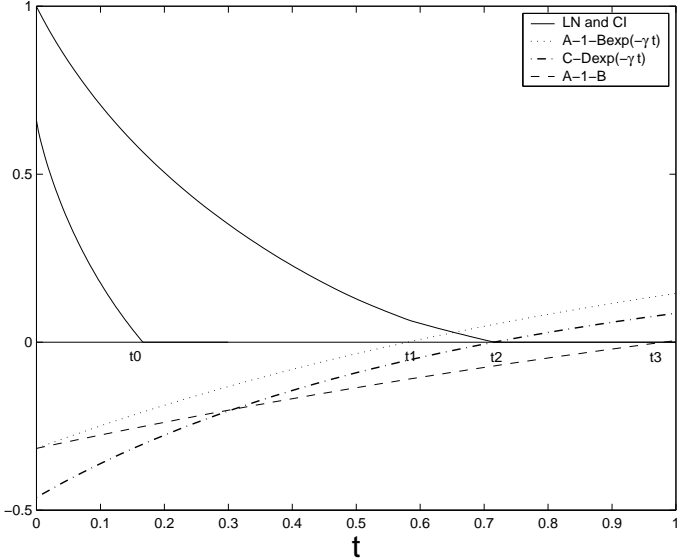}
\caption{The upper solid line is for logarithmic negativity (LN), the down
solid line is for coherent information (CI). The parameters are $\Gamma
=\gamma =\overline{n}=r=0.5$.}
\end{figure}

We have investigated the symmetric damping setting of the two mode squeezed
thermal state, that is, the two modes undergo the same damping and noise.
The generalization to asymmetric damping setting and $x-p$ symmetric
Gaussian state is straightforward with the method developed here. Denote $%
C_i=1-\frac{A_i}{A_1A_2-\left| B\right| ^2},$ $D=\frac B{A_1A_2-\left|
B\right| ^2},$ the $M^{(m)}$ matrix will be
\begin{equation}
M_{ln}^{(m)}=(A_1A_2-B^2)^{-1}C_2^m\sum_k\binom{m-n}{l-k}\binom nk\left(
\sqrt{\frac{C_1}{C_2}}\right) ^{l+n}\left( \frac{\left| D\right| }{\sqrt{%
C_1C_2}}\right) ^{l+n-2k}e^{-\gamma t(n-l)^2},  \label{wave6}
\end{equation}
The whole issue of the positivity of the asymmetric setting is equivalent
that of symmetric setting and omitted here. The necessary and sufficient
criterion of the PPT separability of the state will be

\begin{equation}
\sqrt{C_1C_2}\geq \left| D\right| e^{-\overline{\gamma }t},  \label{wave8}
\end{equation}
and the PPT sufficient criterion is again obtained at the sense of algebraic
programming and perturbation theory. Ineq.(\ref{wave1}) will be generalized
to
\begin{equation}
\sqrt{(A_1-1)(A_2-1)}\geq \left| B\right| e^{-\overline{\gamma }t},
\label{wave3}
\end{equation}

\section{Separability}

We will prove that
\begin{equation}
\text{ }\sqrt{(A_1-1)(A_2-1)}\geq \left| B\right| \Leftrightarrow \text{ }%
\sqrt{C_1C_2}\geq \left| D\right|  \label{wave4}
\end{equation}
is the sufficient condition of the separability of the damped state $\rho $.
Let us first consider a Gaussain density operator $\rho _G$ with its
characteristic function $\chi _G=\exp [-A_1^{\prime }\left| \mu _1\right|
^2-A_2^{\prime }\left| \mu _2\right| ^2+B\mu _1\mu _2e^{ix}+B^{*}\mu
_1^{*}\mu _2^{*}e^{-ix})]$. The Fourier transformation of $\chi _G\exp
(\frac 12\left| \mu \right| ^2)$ is a probability distribution function
(pdf) if $\sqrt{(A_1-1)(A_2-1)}\geq \left| B\right| $\cite{Duan}, where $x$
is absorbed into $\mu $.This pdf enables the P-representation of $\rho _G.$
Hence $\rho _G$ is separable when $\sqrt{C_1C_2}\geq \left| D\right| $. The
P-representation of $\rho $ is a positive integral of the P-representation
of $\rho _G$. Thus $\rho $ is separable. From physical consideration, we may
think $\rho $ is the phase damping of $\rho _G$, thus when $\rho _G$ is
separable, $\rho $ should be separable.

The problem left is that when $\sqrt{C_1C_2}<\left| D\right| $ and $\sqrt{%
C_1C_2}\geq \left| D\right| e^{-\gamma t}$, the state is separable or not.
We have strong evidence to elucidate that the state is not separable,
although we do not give a full proof. The evidence is like this: the state
can not be expressed in P-representation for $\sqrt{C_1C_2}<\left| D\right| $
, $\rho =\int P(\alpha _1,\alpha _2)\left| \alpha _1\alpha _2\right\rangle
\left\langle \alpha _1\alpha _2\right| $ $d^2\alpha _1d^2\alpha _2/\pi ^2$
is only possible for $\sqrt{C_1C_2}\geq \left| D\right| $ $,$ where $%
P(\alpha _1,\alpha _2)$ is a pdf and $\left| \alpha _1\alpha _2\right\rangle
$ denotes two-mode coherent state.

The Fourier transformation will be $P(\alpha )=\int \chi (\mu )\exp [\frac
12\left| \mu \right| ^2-\mu \alpha ^{*}+\mu ^{*}\alpha ]d^4\mu /\pi ^2.$
After the integral of $\mu $, we have
\begin{equation}
P(\alpha )=c\int_x\exp [-E_2\left| \alpha _1\right| ^2-E_1\left| \alpha
_2\right| ^2+Fe^{ix}\alpha _1\alpha _2+F^{*}e^{-ix}\alpha _1^{*}\alpha
_2^{*}],
\end{equation}
where $c=\frac 1{(A_1-1)(A_2-1)-\left| B\right| ^2}$ $E_i=\frac{A_i-1}{%
(A_1-1)(A_2-1)-\left| B\right| ^2},F=\frac B{(A_1-1)(A_2-1)-\left| B\right|
^2}.$ Denote $\alpha _i=r_ie^{i\theta _i},F=\left| F\right| e^{i\phi }$ and $%
\varphi =\theta _1+\theta _2+\phi ,$ then
\begin{equation}
P(\alpha )=c\exp [-E_2r_1^2-E_1r_2^2]\sum_{n=0}^\infty \frac{(\left|
F\right| r_1r_2)^n}{n!}\sum_{l=0}^n\binom nl\exp [-\overline{\gamma }t\left|
2l-n\right| ^2+i(2l-n)\varphi ].
\end{equation}
Clearly $P(\alpha )$ is real, and $P(\alpha )$ is positive. The positivity
of $P(\alpha )$ is warranted by the fact that when $\sqrt{C_1C_2}\geq \left|
D\right| $ , the state is separable, $P(\alpha )$ is a pdf. If we fix $F$
while decreasing $E_i$ to reach a state with $\sqrt{C_1C_2}<\left| D\right| $
, the sign of $P(\alpha )$ will not change by decreasing $E_i$. Thus $%
P(\alpha )$ is positive even when $\sqrt{C_1C_2}<\left| D\right| .$

What left is the singularity of $P(\alpha ).$ We will prove that $P(\alpha )$
is singular if and only if $\sqrt{C_1C_2}<\left| D\right| .$ The singularity
may appear when $\left| \alpha _i\right| =r_i\rightarrow \infty ,$ $P(\alpha
)\rightarrow \infty .$ The maximum of $P(\alpha )$ reaches when $\varphi =0,$
thus in the following, we set $\varphi =0.$ Let $g(z)=\sum_{n=0}^\infty
\frac{\left( z/2\right) ^n}{n!}\sum_{l=0}^n\binom nl\exp [-\overline{\gamma }%
t\left| 2l-n\right| ^2],$ then we have $g(z)\approx \sum_{n=0}^\infty \frac{%
z^n}{n!\sqrt{1+\overline{\gamma }tn}}$ $\approx 1+\sum_{n=1}^\infty \frac{z^n%
}{n!\sqrt{\overline{\gamma }tn}}>1+\frac 1{\sqrt{\overline{\gamma }t}%
}\sum_{n=1}^\infty \frac{z^n}{(n+1)!}=1+\frac 1{\sqrt{\overline{\gamma }t}%
}\frac 1z(e^z-1-z),$ where we have used DeMoirve-Laplace theorem $\binom
nlp^lq^{n-k}\approx $ $\frac 1{\sqrt{2\pi npq}}\exp [-\frac{(l-np)^2}{2npq}].
$ Thus $1+\frac 1{\sqrt{\overline{\gamma }t}}\frac 1z(e^z-1-z)\lesssim
g(z)\leq e^z,$ $\lim {}_{z\rightarrow \infty }\frac{\ln g(z)}z=1.$ We arrive
at $P(r_1,r_2)\rightarrow c\exp [-E_2r_1^2-E_1r_2^2+2\left| F\right|
r_1r_2]\leq c\exp [-2(\sqrt{E_2E_1}-\left| F\right| )r_1r_2]$ when $%
r_1r_2\rightarrow \infty .$ The non-singularity condition of $P(\alpha )$ is
simply $\sqrt{E_2E_1}\geq \left| F\right| ,$ which is equivalent to Ineq.(%
\ref{wave4}).

We now compare all three conditions of the separability of $\rho .$ If $%
\sqrt{C_1C_2}\geq \left| D\right| ,$ the state $\rho $ is separable,
needlessly to say we have $\sqrt{C_1C_2}\geq \left| D\right| e^{-\overline{%
\gamma }t},$ and $\sqrt{(A_1-1)(A_2-1)}\geq \left| B\right| e^{-\overline{%
\gamma }t}$; if $\sqrt{C_1C_2}<\left| D\right| $ and $\sqrt{C_1C_2}\geq
\left| D\right| e^{-\overline{\gamma }t}$, we have $\sqrt{(A_1-1)(A_2-1)}%
\geq \left| B\right| e^{-\overline{\gamma }t},$hence Ineq.(\ref{wave3}) can
be dropped as a necessary condition because it is weak than Ineq.(\ref{wave8}%
), at this case we do not know the state $\rho $ is separable or not, we
suspect that the state is bound entangled; if $\sqrt{C_1C_2}<\left| D\right|
e^{-\overline{\gamma }t}$, the state $\rho $ is entangled. The conditions
are expressed with the curves $A-1-Be^{-\gamma t},C-De^{-\gamma t}$ and $%
A-B-1$ in the figure for the special case of $A_1=A_2,B=B^{*}$. The zero
points of the curves are $t_1,t_2,t_3$, and $t_1\leq t_2\leq t_3$. For
channel without phase damping, the state is a gaussian state. All three
conditions will be the same, the zero points of the curves will coincide and
$t_1=t_2=t_3$.

\section{Conclusions and Discussions}

In conclusion, the phase damping equation of a state is obtained in the form
of characteristic function. It turns out to be a usual dissipation equation
with respect to the phase angle of the complex variable of the
characteristic function. The time evolution solution is given for any
continuous variable state undergo simultaneous amplitude and phase damping
and thermal noise. Two of the criteria are given for the amplitude and phase
damping of a two mode $x-p$ symmetric Gaussian state. One is the sufficient
condition of the damped state. The other is Peres-Horodecki criterion which
is not only necessary but also proved to be PPT sufficient. The proof is at
the sense of algebraic programming and also perturbation theory. The
logarithmic negativity and coherent information of the damped state are
investigated.

The evolution of the state is like this: the entanglement of the state (if
the state is prepared entangled initially) decreases with time, at some time
it reaches $0,$ this time is determined by Peres-Horodecki criterion. Then
the state may be bound entangled at the next time interval, we proved that
the state has not a P-representation at this time interval. The end of this
time interval is the time determined by the sufficient condition of the
separability. After this time the state is separable.

For a channel without phase damping, the state remains a Gaussian state. The
two criteria will coincide\cite{Duan}\cite{Simon}. For pure phase damping
channel, $\Gamma =0,$ $\overline{n}=0,$ we can see that the initially
two-mode vacuum state will never evolve to a separable state. Hiroshima
mentioned this result with numerical calculation\cite{Hiroshima}.

This work was supported by the National Natural Science Foundation of China
(under Grant No. 10347119), Zhejiang Province Natural Science Foundation
(under Grant No. R104265) and AQSIQ of China (under Grant No. 2004QK38)

\section{Appendix}

From the eigenequation of partial transposed density matrix $\rho
^{PT}(t)\left| \Phi \right\rangle =\lambda \left| \Phi \right\rangle ,$ the
eigenequation for characteristic function of it can be deduced as $\int
\frac{d^4\mu }{\pi ^2}\frac{d^4\alpha }{\pi ^2}\chi ^{PT}(\mu ,\mu
^{*},t)\left\langle \beta \right| D(-\mu )\left| \alpha \right\rangle
\left\langle \alpha \right| \left. \Phi \right\rangle =\lambda \left\langle
\beta \right| \left. \Phi \right\rangle .$ Let $\left\langle \alpha \right|
\left. \Phi \right\rangle =\exp (-\frac 12\left| \alpha \right|
^2)\sum_{n=0}^mc_n^{(m)}\alpha _1^{*m-n}\alpha _2^{*n}$, then
\begin{eqnarray*}
I_{mn} &=&\exp (\frac 12\left| \beta \right| ^2)\int \frac{d^4\mu }{\pi ^2}%
\frac{d^4\alpha }{\pi ^2}\chi ^{PT}(\mu ,\mu ^{*},t)\left\langle \beta
\right| D(-\mu )\left| \alpha \right\rangle \exp (-\frac 12\left| \alpha
\right| ^2)\alpha _1^{*m-n}\alpha _2^{*n} \\
&=&\int_x\int \frac{d^4\mu }{\pi ^2}\exp [-A_1\left| \mu _1\right|
^2-A_2\left| \mu _2\right| ^2-B\mu _1\mu _2^{*}e^{ix}-B^{*}\mu _1^{*}\mu
_2e^{-ix}+\mu \beta ^{*}] \\
&&*(\beta _1^{*}-\mu _1^{*})^{m-n}(\beta _2^{*}-\mu _2^{*})^n.
\end{eqnarray*}
Where the integral formula
\[
\int \frac{d^2\tau }\pi \exp [-\left| \tau \right| ^2+\tau \sigma ]f\left(
\tau ^{*}\right) =f\left( \sigma \right)
\]
is used to integrate $\alpha .$ This formula can further be used to
integrate $\mu .$ After the integral of $\mu _1$ we have
\begin{eqnarray*}
I_{mn} &=&\frac 1{A_1}\int_x\int \frac{d^2\mu _2}\pi \exp [-A_2\left| \mu
_2\right| ^2-\frac 1{A_1}B^{*}\mu _2e^{-ix}(\beta _1^{*}-B\mu
_2^{*}e^{ix})+\mu _2\beta _2^{*}] \\
&&*[\beta _1^{*}-\frac 1{A_1}(\beta _1^{*}-B\mu _2^{*}e^{ix})]^{m-n}(\beta
_2^{*}-\mu _2^{*})^n.
\end{eqnarray*}
After the integral of $\mu _2$ we have
\[
I_{mn}=\frac 1{A_1A_2-\left| B\right| ^2}\int_x\left( C_2\beta
_1^{*}+De^{ix}\beta _2^{*}\right) ^{m-n}\left( C_1\beta
_2^{*}+D^{*}e^{-ix}\beta _1^{*}\right) ^n,
\]
with $C_i=1-\frac{A_i}{A_1A_2-\left| B\right| ^2}$ and $D=\frac
B{A_1A_2-\left| B\right| ^2}$. For each $m$ we expand the binomial and
complete the integral of $x,$ $I_{mn}$ will be a polynomial of $\beta _1^{*}
$ and $\beta _2^{*}$. The eigenequation will be
\[
\sum_{n=0}^mc_n^{(m)}I_{mn}(\beta _1^{*},\beta _2^{*})=\lambda
\sum_{k=0}^mc_k^{(m)}\beta _1^{*m-k}\beta _2^{*k}.
\]
By comparing the power of $\beta ^{*},$ and absorbing the phase of $D$ into
the coefficient $c_n^{(m)},$ we at last get a matrix $M^{(m)}$ (as in Eq.(%
\ref{wave6}) ) whose eigenvalues are that of $\rho ^{PT}(t).$ Eq.(\ref{wave7}%
) can be deduced in the same way.

\end{document}